# Recent measurements of proton electric form factor and antiproton-proton elastic scattering

Mohammad Saleem[1], Muhammad Ali[1], Shaukat Ali[2] and Atif Shahbaz[2]
1. Theory Group, Department of Space Science, Punjab University, Lahore. Pakistan.
2. Theory Group, Centre for High Energy Physics, Punjab University, Lahore. Pakistan.

**Abstract**

The Chou-Yang model based on multiple diffraction theory has already been used by us to predict the characteristics of antiproton-proton elastic scattering at very high energies. But the recent precise data on proton electric form factor obtained at JLab essentially deviates from the dipole approximation. We have parametrized the data for the form factor. The computations have been made by using the new expression for the form factor. The implications of the new form factor on the characteristics of antiproton-proton elastic scattering at very high energy have been studied.

The measurements of various characteristics of proton-proton and proton-antiproton elastic scatterings [1-23] have played an important role in the development of high energy physics. The various models have been proposed at various stages [24-39] and modified or demolished by the new experimental results. One of models that has persisted with time is based on the diffraction theory and is called the generalized Chou-Yang model. We would give a physical picture of proton-antiproton elastic scattering in this model. The colliding particles are clusters of objects called partons that collide with each other in pairs. At high energy, when the wavelengths are very small, it is feasible to describe the interaction between these clusters of partons as involving successive collisions between every parton of one cluster with individual partons making up the other cluster. This interaction depends critically upon the relative positions of these partons, and, therefore, upon the pair and possibly higher correlation functions. The multiple scattering effect is small compared to single scattering at small momentum transfer as the peripheral partons dominating the process in this region are not expected to suffer more than one collision before they leave the scattering region. The differential cross section in this case is therefore a consequence of a single scattering. Since the pristine Chou-Yang model tacitly assumes such a scattering, it becomes consistent with the generalized Chou-Yang model and also with the experimental data in the diffraction peak region. At high momentum transfer, that is, at large ! t, where the form factor and, therefore, the single scattering cross section is small, the correlation effect becomes observable. It is this effect that is responsible for the difference between the experiment and the predictions of the pristine Chou-Yang model in the large $-t$ region. In fact, for large $-t$, central partons constituting a colliding particle suffer successive collisions with two or more partons of other particles before leaving the scattering region. This introduces an additional function in the integrand in the expression for the opacity. This must be a slowly varying function of t so that for small angle scattering, that is, for the diffraction peak region, it reduces to ≃ 1, yielding the pristine Chou-Yang model. For large angle scattering, i.e., beyond the diffraction peak region, this function makes significant contribution; this is why in this region only the multiple diffraction theory that takes into account the role of this function, is successful.

The model depends upon two parameters, K and α, to be determined by experimental data. Recently, Saleem et al. [40] have proposed a formula that at any given high energy yields the values of these parameters. By using the anisotropy function, $(1 + At)/(1 - At)^n$, $A = 1/3$ GeV$^{-2}$, $n = 1/3$, the values of differential and total cross sections and other characteristics of antiproton-proton elastic scattering were obtained and found to be consistent with experiment. However, recently precise measurements of proton form factor have shown [41] that for $Q^2 = -3.5$ GeV$^2$, the shape is significantly different from the dipole approximation or other forms based on earlier measurements in the region where precise measurements have been made. We have found that in view of the recent precise measurements, the expression for the proton electric form factor has to be changed from

$$G_P = 0.6405 \exp(4t) + 0.33 \exp(0.85t) + 0.028 \exp(0.22t) + 0.0015 \exp(0.05t)$$

to

$$G_P = 0.6405 \exp(4t) + 0.33 \exp(0.75t) + 0.023 \exp(0.20t) + 0.0012 \exp(0.04t).$$

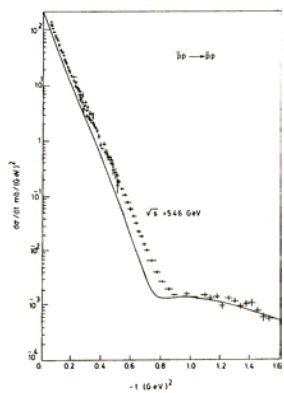  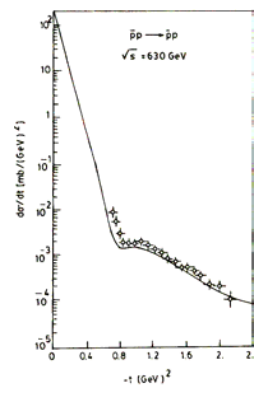

Fig 1                     Fig2

The programme was run after substituting the new expression for the form factor. It was found that for large values of !t, the values of differential cross section decreased significantly. This clearly indicated that the anisotropy function had to be changed. For that purpose, we chose $A = 0.25$ GeV$^{-2}$ and $n = 0.25$ in the expression for the anisotropy function. The results obtained with this choice may be considered consistent with experiment. The results obtained for differential cross sections at $\sqrt{s} = $ 546, 630, 900, 1020, 1080 GeV and 10 TeV, along with the experimental data wherever available, are shown in Figs. 1 to 4.

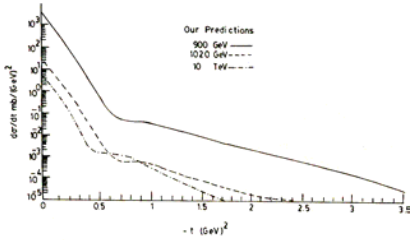

Fig 3

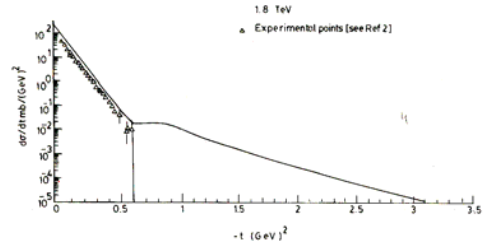

Fig 4

We conclude that the new precise measurements of the proton electric form factor reflect the fact that the anisotropy function for antiproton-proton elastic scattering has to be changed. We may point out that the anisotropy function has been taken as energy independent; once it is determined, the same expression is to be used at all energies.

The total cross section values are 59.5, 64.4, 65.7 and 71.7 mb at the centre of mass energies 546 GeV, 900 GeV, 1.02 TeV and 1.8 TeV, respectively. These are consistent with the experimental values of $61.9 \pm 1.5$, $65.3 \pm 0.7 \pm 1.5$, $61.1 \pm 5.7$ and $72.8 \pm 3.1$ mb.

**Figure Captions**

**Fig.1**. The differential cross section at $\sqrt{s}$ = 546 GeV plotted against ! t. The solid curve represents our prediction. The experimental values are taken from Bozzo et al. [8,10].

**Fig.2**. The differential cross section at $\sqrt{s}$ = 630 GeV plotted against ! t. The solid curve represents our prediction. The experimental values are taken from Bernard et al. [13].

**Fig.3**. Predictions of differential cross section at $\sqrt{s}$ = 900 GeV, 1.02 TeV and 10 TeV plotted against ! t. A $10^{-1}$ factor between curves is omitted. The differential cross section values for 900 GeV have been multiplied by 10.

**Fig.4**. The differential cross section at $\sqrt{s}$ = 1.8 TeV plotted against ! t. The solid curve represents our prediction. The experimental values are taken from Amos et al. [21].